\documentclass[aps,prd,twocolumn,showpacs,floats,floatfix,letterpaper,nofootinbib,superscriptaddress,]{revtex4}

\usepackage{amssymb,amsmath,latexsym,mathrsfs}
\usepackage{graphicx}
\usepackage{epsfig}
\usepackage{multirow}
\usepackage{array}
\usepackage{color}

\newcommand{\omegae}{\Omega_{\textrm{eDE}}}
\newcommand{\cvis}{c^2_{\textrm{vis}}}
\newcommand{\ceff}{c^2_{\textrm{eff}}}

\newcommand{\lambdasf}{\lambda_{21}}
\newcommand{\ubot}{u_\perp}
\newcommand{\kbot}{k_\perp}
\newcommand{\bubot}{{\bf u}_\perp}
\newcommand{\ubotmin}{\ubot^{\min}}

\newcommand{\upara}{u_\parallel}
\newcommand{\kpara}{k_\parallel}

\newcommand{\PTb}{P_{\dTb}}
\newcommand{\dTb}{\delta T_b}
\newcommand{\Dang}{D_\uA}
\newcommand{\uA}{\mathrm{A}}
\newcommand{\bkbot}{\bf {k}_\perp}

\begin{document}

\preprint{IFIC/}

\def\thefootnote{\fnsymbol{footnote}} \title{Current constraints on
  early and stressed dark energy models\\ and future 21 cm perspectives}
\author{Maria Archidiacono} 
\affiliation{Department of Physics and Astronomy, Aarhus University, 8000 Aarhus C, Denmark.}

\author{Laura Lopez-Honorez}
\affiliation{Theoretische Natuurkunde,\\
Vrije Universiteit Brussel and The International Solvay Institutes,\\
Pleinlaan 2, B-1050 Brussels, Belgium.}

\author{Olga Mena} 
\affiliation{Instituto de F\'isica Corpuscular (IFIC), CSIC-Universitat de Valencia, E-46071, Spain.}

\begin{abstract}

{Despite the great progress of current cosmological measurements, the nature of the dominant component of the universe, coined {\it dark energy}, is still an open question. {\it Early Dark Energy} is a possible candidate which may also alleviate some fine tuning issues of the standard paradigm.  Using the latest available cosmological data, we find that the 95\% CL upper bound on the early dark energy density parameter is $\omegae\,<\,0.009$. On the other hand, the dark energy component may be a stressed and inhomogeneous fluid. If this is the case, the effective sound speed and the viscosity parameters are unconstrained by current data. Future omniscope-like $21$~cm surveys, combined with present CMB data, could be able to distinguish between standard quintessence scenarios from other possible models with $2\sigma$ significance, assuming a non-negligible early dark energy contribution. The precision achieved on the $\omegae$ parameter from these $21$~cm probes could be below $\mathcal{O} (10\%)$.}
\end{abstract}

\maketitle

\section{Introduction}
The nature of the mysterious dark energy component that currently
dominates the energy content of the universe reveals new physics
missing from our universe's picture, and constitutes the fundamental
key to understand the fate of the universe.  The most economical
explanation of the dark energy component attributes this energy density to the
one of the vacuum, i.e., a cosmological constant scenario. Together
with cold dark matter (CDM), the so-called $\Lambda$CDM scenario can
account for present data with a flat universe made up of roughly
$30\%$ dark matter and $70\%$ dark energy. In this minimal model, the
dark energy equation of state, $w$, which corresponds to the ratio of
the dark energy pressure to the dark energy density, is constant and equal to 
$-1$. However, this simple picture suffers from severe fine tuning
theoretical issues (see Ref.~\cite{Copeland:2006wr} and
references therein) as well as from problems with observations related
to the matter power spectrum on scales of a few Mpc and
below~\cite{Moore:1999gc,Bode:2000gq,Penarrubia:2012bb,BoylanKolchin:2011dk,Ferrero:2011au,Weinberg:2013aya}. Possible
alternatives to alleviate them have been extensively explored. Perfect
dark energy fluids, characterised either by a constant ($w\neq-1$) or
by a time varying dark energy equation of state $w(a(t))$, or scalar
field models, are the most popular options considered in the
cosmological data analyses, as their parameterizations require few
extra parameters (two at most) to be added to the usual $\Lambda$CDM scenario.

There exists also alternative scenarios, in which the gravitational
sector is modified, leading to a modification of Einstein's equations
of gravity on large scales. Modifications of gravity (see
e.g. \cite{DeFelice:2010aj} and references therein) incorporate models
with extra spatial dimensions or an action which is non-linear in the Ricci scalar. There are also non-perfect fluid models, as
Chaplygin gas cosmologies~\cite{Bento:2002ps}, which involve more
parameters than just one equation of state $w$. Of particular interest here
is the {\it Early Dark Energy} (hereafter EDE) case, as it arises as a
natural hypothesis of dark
energy~\cite{Wetterich:2004pv,Doran:2006kp,odea,Calabrese:2010uf}. EDE
differs from the cosmological constant because it is not negligible in
the early universe and the contribution depends on the initial density
parameter $\omegae$. Furthermore, the EDE model considered here is
based on a generic dark energy fluid which is inhomogeneous.  Density
and pressure are time varying, therefore the equation of state is not
constant in time.  The phenomenological analyses of these inhomogeneous dark
energy models usually require additional dark energy clustering
parameters, i.e. the dark energy effective sound speed and the dark
energy anisotropic stress.  The sound speed $\ceff$
\cite{Hu:1998kj,Hu:1998tk,ceff} is defined as the ratio between the
dark energy pressure perturbation and the dark energy density contrast in the rest
frame of the fluid, $\ceff\equiv(\delta P/\delta \rho)_{\rm rest}$.
In the simplest quintessence models, $\ceff=1$, while the
anisotropic stress is zero. The effective sound speed determines the
clustering properties of dark energy and consequently it affects the
growth of matter density fluctuations. Therefore, in principle, its presence could be revealed in large scale structure observations. The growth of
perturbations can also be affected by the anisotropic stress
contributions~\cite{Hu:1998kj,Hu:1998tk,cvis} which lead to a
damping in the velocity perturbations. In the parametrization used
here, the damping effect is driven by the viscosity parameter $\cvis$
which links the anisotropic stress to the velocity perturbation and
the metric shear.

Despite the precision achieved by the combination of Cosmic Microwave
Background (CMB) measurements from the Planck satellite~\cite{planck},
Baryon Acoustic Oscillation (BAO) data from a number of galaxy
surveys~\cite{Anderson:2013zyy,Beutler:2011hx,Busca:2012bu,Kirkby:2013fh,Slosar:2013fi}
and Supernovae Ia luminosity distance
measurements~\cite{Suzuki:2011hu} in the extraction of the dark energy
equation of state parameter, $w=-1.06\pm 0.06$ at
$68\%$~CL~\cite{Anderson:2013zyy}, the nature of the dark energy
component remains unknown. Therefore, it is mandatory to carefully study
other possibilities including the one of an EDE
component, as well as the clustering properties of the dark energy
fluid.  In this paper we shall address both issues, relaxing the
perfect fluid assumption and considering current cosmological data, in
addition to the recent BICEP2 measurements of the B-modes power spectrum~\cite{Ade:2014xna} .

We also explore the possibility of constraining an EDE component
and/or a stressed dark energy fluid with future $21$ cm surveys. The
next generation of radio experiments, which will image the neutral
intergalactic medium (IGM) in $21$~cm emission/absorption, will
provide a unique probe of the universe at higher redshifts ($z>6$)
which lie out of the reach of galaxy surveys and CMB experiments. The
$21$~cm line signal presents several advantages compared to
traditional cosmic and astrophysical probes, see
e.g.~\cite{Loeb:2003ya}, and it could be used to test the nature of dark
energy~\cite{Wyithe:2007rq}. The future
generation of radio interferometers testing the $21$~cm signal,
including the Squared Kilometer Array (SKA)~\cite{Mellema:2012ht} and
omniscopes~\cite{Tegmark:2008au,Zheng:2013tpz}, may provide extra
constraints on the cosmological parameters probing the Epoch of
Reionisation (EoR) or the high redshift window, see
e.g.~\cite{Mao:2008ug,Clesse:2012th}. In addition, the $21$~cm signal
can also be used at low redshifts ($z<5$), offering a
competitive cosmological probe for unraveling the nature of the
component responsible for the present universe's accelerated
expansion~\cite{Chang:2007xk, Hall:2012wd}.

The structure of the paper is as follows.  Sections \ref{sec:ede} and
\ref{sec:stress} describe the early and stressed dark energy models
evolution in terms of the background and perturbation variables. In
Sec.~\ref{sec:methodanddata} we present the method and data followed
in the numerical analyses presented in
Sec.~\ref{sec:current}. Section~\ref{sec:future} addresses the future
perspective and constraints from $21$~cm surveys by means of a Fisher
matrix forecast analysis. Finally, we draw our conclusions in
Sec.~\ref{sec:conclusions}.

\section{Early Dark energy models}
\label{sec:ede}

The concept of EDE cosmology was introduced in~\cite{Wetterich:2004pv}
and studied in several subsequent works following different
possible effective parametrizations of the evolution of the dark
energy fluid, see
e.g.~\cite{Doran:2006kp,Calabrese:2010uf,Pettorino:2013ia,planck}. Here
we follow Ref.~\cite{Doran:2006kp} to describe the evolution of the
background dark energy density from the high redshift, constant value
$\omegae$ until its present-day value $\Omega_{\rm DE}^0$ (assuming a
flat universe with $\Omega_{\rm DE}^0+\Omega_{\rm m}^0=1$):
\begin{equation}
\Omega_{\rm DE}(a) =\frac{\Omega_{\rm DE}^0 - \omegae \left(1- a^{-3 w_0}\right) }{\Omega_{\rm DE}^0 + \Omega_{m}^{0} a^{3w_0}} + \omegae \left (1- a^{-3 w_0}\right).
\label{eq:odea}
\end{equation}
The evolution of $w(a)$ in this EDE parametrization reads
\begin{equation}
w(a) = -\frac{1}{3[1-\Omega_{\rm DE}(a)]} \frac{d\ln \Omega_{\rm DE}(a)}{d\ln a} + \frac{a_{eq}}{3(a + a_{eq})},
\label{eq:wa}
\end{equation}
where $a_{eq}$ is the scale factor at matter-radiation equality era.  The
time dependent equation of state $w(a)$ typically traces the dominant
component of the universe at each epoch: first $w\simeq1/3$ during the
radiation dominated period, then $w\simeq 0$ during the matter
dominated era and finally $w\rightarrow w_0$ in the present
epoch. The current value of the equation of state parameter $w_0$
might be different\footnote{Notice that the clustering
  properties of a universe with $-1<w<-1/3$ deviate from those of a
  $\Lambda$CDM universe with $w=-1$ and therefore it can be inconsistent with
  observations~\cite{effects}.} from $-1$.

\section{Stressed Dark energy models}
\label{sec:stress}

Using the notation of Ref.~\cite{Ma} and assuming the synchronous
gauge, we follow~\cite{Calabrese:2010uf} to describe the dark energy
scalar perturbation evolution equations in Fourier space for the
density contrast ($\delta$), the velocity divergence ($\theta$) and
the anisotropic stress perturbation ($\sigma$):
\begin{eqnarray}
\frac{\dot{\delta}}{1+w}&=& 
-\left[k^{2}+9\left(\frac{\dot{a}}{a}\right)^{2}\left(\ceff-w+\frac{\dot{w}}{3(1+w)(\dot{a}/a)}\right)\right]\frac{\theta}{k^{2}} \nonumber\\ 
&&-\frac{\dot{h}}{2}-3\frac{\dot{a}}{a}(\ceff-w)\frac{\delta}{1+w}; \\
\label{eq:pertdelta}
\dot{\theta}&=&-\frac{\dot{a}}{a}(1-3\ceff)\,\theta+
\frac{\delta}{1+w}\ceff k^{2}-k^{2}\sigma;\\
\label{eq:perttheta}
\dot{\sigma}&=&-3\frac{\dot{a}}{a}\left[1-\frac{\dot{w}}{3w(1+w)(\dot a/a)}\right]\,\sigma\nonumber\\
& & +\frac{8\cvis}{3(1+w)}\left[\theta+\frac{\dot{h}}{2}+3\dot{\eta}\right]~,
\label{eq:pertsigma}
\end{eqnarray}
where $\ceff$ denotes the effective sound speed. In the last equation, the velocity and the metric shear (sometimes
referred to as $H_T= -(h/2 + 3 \eta)$) are related to the dark energy
shear stress through the viscosity parameter $\cvis$. The latter
relation was first introduced in Ref.~\cite{Hu:1998kj}~\footnote{Note that $\sigma$ here is related to the variable $ \pi$
  in~\cite{Hu:1998tk} through the relation $\sigma=(2/3)\,
  \pi/(1+w)$.} and relates directly the anisotropic stress with the damping of velocity fluctuations on shear-free frames ($H_T$ = 0), 
if $\cvis> 0$. We have also addressed the contribution of the dark energy
shear stress  to the evolution equations for the tensor perturbations.

The differential equations above govern the clustering properties of the
dark energy fluid, and we shall solve them and compare the results to
current and future observations using the methods detailed in the
following sections.

\section{Method and data for current constraints}
\label{sec:methodanddata}
We have modified the latest version of the Boltzmann equations solver
\textrm{CAMB} \cite{camb} in order to account for Eqs.~(\ref{eq:odea})-(\ref{eq:pertsigma}).

The parameter space contains the six standard parameters of the
$\Lambda$CDM model
\begin{equation}
\{\Omega_{\rm b}h^2,\,\Omega_{\rm c}h^2,\,\theta,\,\tau,\,n_{\rm s},\,\ln{(10^{10} A_{\rm s})}\},
\end{equation}
where $\Omega_{\rm b}h^2=\omega_{\rm b}$ is the present physical
energy density in baryons, $\Omega_{\rm c}h^2=\omega_{\rm c}$ is the
present physical cold dark matter energy density, $\theta$ is the
angular scale of the sound horizon, $\tau$ is optical
depth to reionisation and $n_{\rm s}$ and $A_{\rm s}$ are the spectral
index and the amplitude of primordial scalar perturbations at a pivot
scale $k=0.05\,{\rm Mpc}^{-1}$, respectively. 

Since we include tensor perturbations, we have also considered the tensor-to-scalar ratio $r$ parameter, defined relatively to the same
pivot scale of the scalar perturbations, $k=0.05\,{\rm
  Mpc}^{-1}$. Finally, we include all the parameters describing the
EDE model evolution (see Secs.~\ref{sec:ede} and \ref{sec:stress}):
\begin{equation}
\{\omegae,\,w_0,\,\cvis,\,\ceff\}.
\end{equation}
We assume flat priors on the parameters as listed in Tab.~\ref{tab:priors}. The sampling of the parameter space is performed through the Monte Carlo Markov Chain (MCMC) public package  \textrm{CosmoMC} \cite{cosmomc}.

\begin{table}[h!]
\begin{center}
\begin{tabular}{c|c}
\hline\hline 
 Parameter & Prior\\
\hline
$\Omega_{\rm b}h^2$ & $0.005 \to 0.1$\\
$\Omega_{\rm c}h^2$ & $0.01 \to 0.99$\\
$\theta$ & $0.5 \to 10$\\
$\tau$ & $0.01 \to 0.8$\\
$n_{\rm s}$ & $0.5 \to 1.5$\\
$\ln{(10^{10} A_{s})}$ & $2.7 \to 4$\\
$r$ & $0\to1$\\
\hline
$\omegae$ & $0 \to 0.1$\\
$w_0$ &  $-1 \to 0$\\
$\ceff$ & $0\to1$\\
$\cvis$ & $0\to1$\\
\hline\hline
\end{tabular}
\caption{Range of the flat priors for the cosmological parameters considered here.}
\label{tab:priors}
\end{center}
\end{table}

The Bayesian inference is based on the CMB temperature anisotropy power spectrum of the Planck experiment, implemented following the prescriptions of Ref.~\cite{Ade:2013kta}. We have also considered the CMB polarization measurements from the nine-year data release of the WMAP satellite~\cite{Bennett:2012zja}. In the following, we shall refer to the former data as WP.  The maximum multipole number of  the Planck temperature power spectra is $\ell_{\rm max}=2500$. The WP measurements reach a maximum multipole $\ell=23$, see Ref.~\cite{Bennett:2012zja}. In order to directly constrain the tensor-to-scalar ratio $r$, the nine-bins measurements of the B-modes polarization power spectrum from the BICEP-2 collaboration\cite{Ade:2014xna} are included.

\section{Current cosmological constraints}
\label{sec:current}
In this section we apply the data sets described above, using the MCMC method, to four possible scenarios:
\begin{itemize}
\item {\bf Case 1}: In this scenario, both the early dark energy
  component $\omegae$ and the dark energy perturbation parameters
  $\ceff$ and $\cvis$ are free parameters, with the priors specified
  in Tab.~\ref{tab:priors}. We also consider in this case the current value of the dark energy equation of state, $w_0$, see Eq.~(\ref{eq:odea}), as a free parameter. 
\label{case1}
\item {\bf Case 2}: The early dark energy component $\omegae$ and $w_0$ are
  free parameters, but the dark energy perturbations are fixed to their
  standard values: $\ceff=1$ and $\cvis=0$ (i.e. no anisotropic stress
  contribution is considered in this case).
\label{case2}
\item {\bf Case 3}: We consider no early dark energy component
  ($\omegae=0$) but the dark energy perturbations $\ceff$
  and $\cvis$ are both free parameters, varying with a flat prior in the
  range $[0,1]$, as well as a constant dark energy equation of state $w$, which varies with a prior in the range $[-1,0]$.
\label{case3}
\item {\bf Case 4}: We consider a simple $w$CDM cosmology, i.e., a
  cosmological scenario with a constant dark energy equation
  of state, which is allowed to freely vary in the range $[-1,0]$.
\label{case4}
\end{itemize}

\begin{table*}
\begin{center}
\begin{tabular}{lll}
\hline \hline
  &Planck+ WP & Planck +WP + BICEP-2\\
\hline
\hline
Case 1& & \\
\hline
${\omegae}$   &  $<0.015$& $<0.010$\\
$w_0$  & $<-0.658$& $<-0.722$\\
$r$ & $<0.09$ & $0.15\pm0.04$ \\
$n_{\rm s}$ & $0.960\pm0.008$ & $0.963\pm0.007$ \\
\hline
Case 2& & \\
\hline
${\omegae}(\ceff=1, \cvis=0)$   &  $<0.012$& $<0.009$\\
$w_0(\ceff=1, \cvis=0)$  & $<-0.659$& $<-0.722$\\
$r$ & $<0.10$ & $0.16\pm0.04$ \\
$n_{\rm s}$ & $0.960\pm0.007$ & $0.963\pm0.008$ \\
\hline
Case 3& & \\
\hline
${\omegae}$   &  $0$& $0$\\
$w$  & $<-0.647$ &$<-0.709$\\
$r$ & $<0.11$ & $0.16\pm0.04$ \\
$n_{\rm s}$ & $0.960\pm0.007$ & $0.964\pm0.007$ \\
\hline
Case 4& & \\
\hline
${\omegae}(\ceff=1, \cvis=0)$   &  $0$& $0$\\
$w(\ceff=1, \cvis=0)$  & $<-0.655$& $<-0.705$\\
$r$ & $<0.11$ & $0.16\pm0.04$ \\
$n_{\rm s}$ & $0.960\pm0.008$ & $0.964\pm0.007$ \\
\hline \hline
\end{tabular}
\caption{Mean values with $1\sigma$ errors and $2\sigma$ upper bounds
  for the $\omegae$ parameter as well as for the most correlated cosmological
  parameters for the different possible cases described in
  Sec.~\ref{sec:current}. The dark energy perturbation parameters
  $\ceff$ and $\cvis$ are not listed in this table, as current cosmological data 
  are unable to constrain them.}
\label{tab:ede_finaltest}
\end{center}
\end{table*}

Table~\ref{tab:ede_finaltest} shows the mean values with $1\sigma$
errors and the $2\sigma$ upper bounds for the EDE parameters following
the case order listed above. Notice first that we do not show the values for the dark energy
  perturbation parameters ($\ceff$ and $\cvis$), since current CMB measurements are unable to constrain them. 
  Secondly, when setting $\ceff=1$ and
  $\cvis=0$ (see Case 2 above), we find an upper limit on the early dark
energy parameter $\omegae< 0.012$ at $95\%$~CL. The former bound is
looser than the one reported by the Planck collaboration, $\omegae<
0.010$ at $95\%$~CL with the same data sets (Planck temperature and WP
data). The larger value that we get on $\omegae$ is related to the
degeneracy between this parameter and the tensor-to-scalar ratio $r$,
as we shall explain below. The addition of the BICEP2 data makes our
$95\%$~CL upper limit on $\omegae$ tighter ($\omegae<
0.009$ at $95\%$~CL).  When allowing the dark
energy perturbations $\ceff$ and $\cvis$ to be free parameters
(Case 1 above), the $95\%$~CL upper bound $\omegae$ degrades but not
 significantly: we find $\omegae< 0.015$ ($\omegae< 0.010$)
at $95\%$~CL before (after) combining Planck and WP measurements with
BICEP2 data.

In general, the results for the standard $\Lambda$CDM cosmological parameters do not deviate significantly from their expected mean values and errors.  This can be noticed by comparing the first three cases depicted in Tab.~\ref{tab:ede_finaltest} with the last rows, which show the expectations within the $w$CDM cosmological scenario. Indeed, the current value of the dark energy equation of state $w_0$ does not show a very strong dependence on the dark energy perturbation parameters, as its $95\%$~CL upper bound remains unaffected when  $\ceff$ and $\cvis$ are both freely varying. Concerning the value of $n_s$, its mean value is strongly affected when including BICEP2 data in our numerical analyses, regardless of the dark energy scenario.

\begin{figure*}

\includegraphics[width=18cm]{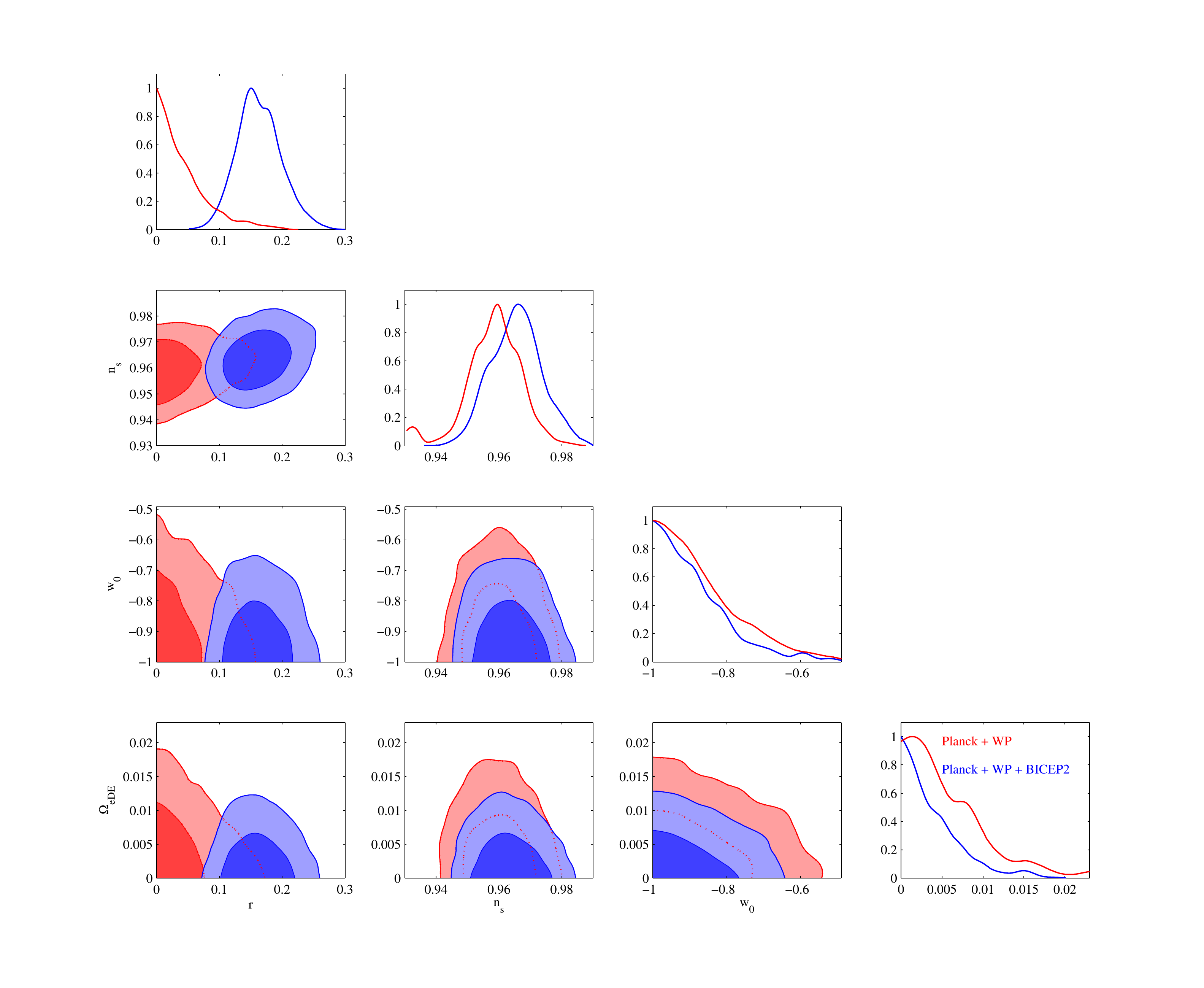}
 \caption{{\it 2D plots}: Red (Blue) contours show the $68\%$ and $95\%$~CL allowed
  regions from Planck + WP (Planck + WP + BICEP2). {\it 1D plots}: Red
  (Blue) lines depict the marginalised one-dimensional posteriors from
  Planck + WP (Planck + WP+ BICEP2) measurements. In this case, both
  the dark energy perturbation parameters and the EDE component are
  free parameters (see Case 1 of Tab.~\ref{tab:ede_finaltest}). BICEP2
  measurements point towards a non-zero value of $r$. As a
  consequence, since $\omegae$ and $r$ are anti-correlated, the
  constraints on $\omegae$ are tighter when considering BICEP2 data in
  the numerical analyses (see the results depicted in
  Tab.~\ref{tab:ede_finaltest}).}

\label{fig:ede_finaltest}
\end{figure*}

Figure~\ref{ede_finaltest} shows the marginalised 2D plots and the
posteriors involving the most relevant cosmological parameters here in the case in which both the early dark energy
component $\omegae$ and the perturbation parameters $\ceff$ and
$\cvis$ are allowed to vary freely (see Case 1 above). The red
contours refer to the results arising from the analysis of Planck + WP
data, while the blue contours include BICEP2 as well. The marginalised
2D plot in the bottom left corner, in the ($\omegae$, $r$)
plane, shows the degeneracy between the EDE component and the tensor-to-scalar ratio $r$. There
exists a mild anti-correlation between these two parameters, which can
be easily understood: both parameters show an effect at very large scales,
increasing the power at very low multipoles. As the BICEP2 data constrain
$r$ to be different from zero, the $2\sigma$ upper bound on $\omegae$
is tighter, in order to compensate the contribution from the tensor modes
at large scales.  A similar effect can also be noticed in the 2D
marginalised plot in the ($w_0$, $r$) plane: given the
anti-correlation between $w_0$ and $r$, the BICEP2 measurements of $r$
reduce the upper bound on $w_0$. There also exists a degeneracy between
the $\omegae$ and $w_0$ parameters, as can be noticed from the right
lower panel of Fig.~\ref{fig:ede_finaltest}: larger (smaller) values
of the present dark energy equation of state, $w_0$, allow for smaller
(larger) values of the EDE parameter,
$\omegae$. Therefore, these two parameters are anti-correlated, as can
be learnt from Eq.~(\ref{eq:odea}): for a given value of the $\omegae$
parameter and the scale factor $a$, the quantity $\Omega_{\rm DE}$
grows as the value of $w_0$ does.


\section{$21$~cm Forecasts}
\label{sec:future}

In this section, we follow the description of
  Ref.~\cite{Clesse:2012th} for the 21 cm brightness background
temperature $T_b(z)$, for the evolution equations of the linear perturbation $\delta T_b(z)$ as well as for the reionisation model implementation.

The study of the 21 cm signal requires to deal with the angular
location on the sky plane ${\bf \theta}$, and with the frequency
difference $\Delta f$ of the signal to a central 21 cm line of
redshift $z$.  The dual coordinates of this system are denoted by
${\bf \ubot}$ and $\upara$, and they are related to the standard
comoving wavevector ${\bf k}$ components as follows:
\begin{equation}
{  \bubot = \Dang(z) \bkbot,} \qquad \upara = y(z) \kpara,
\end{equation}
where $\Dang(z)$ is the angular comoving distance and 
 \begin{equation}
  y(z) = \frac{\lambdasf (1+z)^2 }{H(z)}\,,
\end{equation}
where $\lambdasf$ is the 21 cm wavelength (in the rest frame) and $H(z)$
is the Hubble rate.  The 21 cm brightness temperature power spectrum
relevant for our analyses, $\PTb({\bf u})$, is related to $\PTb({\bf
  k})$ as follows:
\begin{equation}
  \PTb({\bf u}) = \frac{\PTb({\bf k})}{\Dang(z)^2 y(z)}.
\end{equation}

For the Fisher matrix analysis, we have adopted the formalism of Refs.~\cite{Mao:2008ug,Clesse:2012th}.  
Assuming that $\PTb({\bf u})$ is gaussian-distributed, we can approximate the Fisher matrix by
\begin{equation}
  F_{ab} = \dfrac{1}{2} \sum_{\upara,\ubot} \frac{N_c}{\left[\PTb({\bf u}) +
      P_{noise}\right]^2}
  \frac{\partial \PTb({\bf u})}{\partial\lambda_a} \dfrac{\partial
    \PTb({\bf u})}{\partial \lambda_b}\,,
  \label{eq:fish}
\end{equation}
where $\lambda_{a,b}$ are the cosmological parameters involved in the Fisher forecast analysis, and
\begin{equation}
  N_c = \frac{4 \pi f_{sky}}{\Theta^2} 2 \pi \kbot \delta \kbot \delta \kpara 
  \dfrac{V}{(2\pi)^3}\,,
\end{equation}
is the number of independent cells probed for a given value of ${\bf
  u}$ (or ${\bf k}$), $V$ is the comoving volume covered and $\Theta$ is
the angular patch in the sky~\footnote{$\Theta$ is taken to be lower than $1$~rad to be in agreement
with the flat-sky approximation.}.  In Eq.~(\ref{eq:fish}), $P_{noise}$ is given by~\cite{Zaldarriaga:2003du,Clesse:2012th}:
\begin{equation}
\label{eq:Pnoiseu}
P_{noise}({\bf u}) \simeq \frac{4 \pi f_{sky}}{\Omega_{\rm fov}} \frac{\lambda^2}{D^2 f_{cover}^2}
  \frac{T_{sys}^2}{B_W t_{obs}} \,,
\end{equation}
with $f_{sky}$ the fraction of the sky covered by the survey, $\Omega_{\rm fov}$ the
field of view, $\lambda$ the redshifted wavelength of the signal,
$T_{sys}$ the system temperature, $D$ the size of the
array, $B_W$ the experiment's bandwidth and $f_{cover}$ the
covering factor of the array. Beam effects at small scales can be
incorporated by multiplying  Eq.~(\ref{eq:Pnoiseu})  by the factor $\exp{[
  \bubot^2/(4\sqrt{\ln 2}/\theta_{fw})^2]}$, 
  with $\theta_{fw}=0.89\lambda/D$, see Ref.~\cite{Clesse:2012th}.

In what follows, we consider two possible $21$~cm experiment configurations. The
first one is a CHIME-like~\cite{Newburgh:2014toa} experiment, covering
a low redshift range $0.8<z<2.5$.  In our analyses we use a setup
similar to the one considered in~\cite{Hall:2012wd}. The second one is
an omniscope-like instrument sensitive to the EoR. In the latter case,
we follow the setup of Ref.~\cite{Clesse:2012th}. In our treatment of
the Fisher matrix, we use a convolution of the signal with the
frequency window function associated with the mean redshift of
observation. This method helps in reducing the degeneracy between the cosmological parameters $\tau$
and $\ln (A_S)$ when considering one single redshift slice for an
omniscope-like experiment~\cite{Clesse:2012th}. Notice that, in what
follows, we shall assume that most of the foregrounds can be
eliminated, assumption which is still under active research (see
e.g.~\cite{forgnd}). We also neglect the fact that ionising sources
could affect the 21cm perturbations, providing extra contributions to
the power spectrum~\cite{Mao:2008ug,Clesse:2012th}. Therefore the analysis presented here should
be regarded as an optimistic appraisal of the 21 cm signal potential to
constrain both an EDE component and its clustering properties.

\begin{figure}[h!]
  \begin{center}
    \hspace{-1cm}
    \includegraphics[width=9cm]{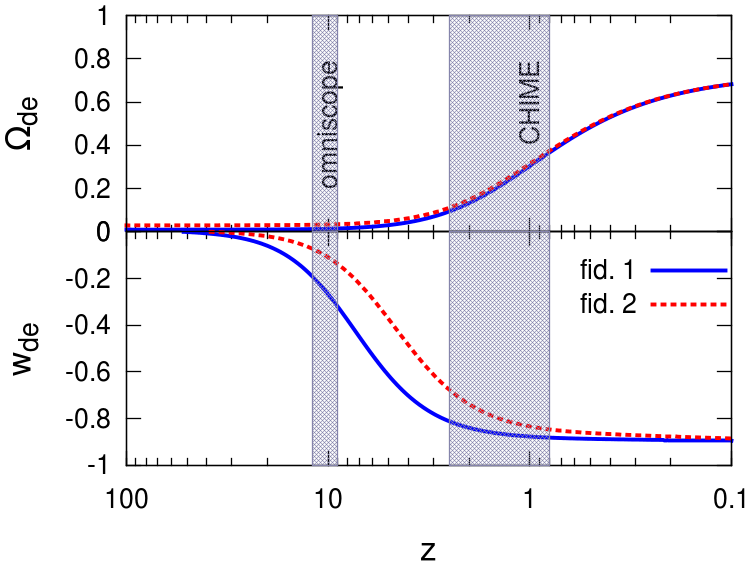}
  \end{center}
  \caption{Evolution of the background quantities for the fiducial
    models of Tabs.~\ref{tab:chime12} and \ref{tab:omn12}. The
    redshift ranges tested by the 21 cm experiments considered here are shown by the
     grey rectangular areas.}
  \label{fig:fid}
\end{figure}

  We present results for two fiducial cosmological models: the
  fiducial model 1 (2) with $\omegae=0.01$ ($0.03$), $\cvis=0$
  ($0.33$) and $\ceff=1$ ($0.33$), both of them assuming the same value for the dark energy equation
  of state at present,  $w_0=-0.9$. Figure~\ref{fig:fid} shows the evolution of the
  background quantities $\Omega_{\rm DE}(z)$ and $w(z)$, see
  Eqs.~(\ref{eq:odea}) and (\ref{eq:wa}), as a function of the
  redshift, for these two possible fiducial cosmologies. The redshift
  ranges tested by the two possible $21$~cm future experiments
  considered here are depicted by the grey rectangular
  zones. Notice that both experiments are located where the difference
  among the expansion histories for these two fiducial models is non-negligible.  Therefore, one would expect to have sensitivity to
  distinguish between different cosmological backgrounds when exploring the $21$~cm power
  spectrum in the two redshift ranges depicted in Fig.~\ref{fig:fid}.

%

\subsection{CHIME $0.8<z<2.5$}

In Tab.~\ref{tab:chime}, we provide the value of the parameters
specifying the CHIME experiment considered in our analyses, which are
similar to those considered in Ref.~\cite{Hall:2012wd}. For the system
temperature, we have taken $T_{sys}= \left[ 40 + 5 \left(\nu /710 {\rm
    Mhz}\right)^{-2.6}\right]$~K, where $\nu$ is the redshifted
frequency of the 21 cm signal. We have also considered a comoving
number density of sources of $0.03h^3$Mpc$^{-3}$, contributing to 
shot-noise. 

\begin{table} [h!]
  \begin{center}
  \begin{tabular}{c c c c c c}
    \hline
    redshift slices&$B_W$& $D$ & $f_{cover}$& $t_{obs}$ &$f_{sky}$\\
    \hline
   0.8/1/2/2.5& 2 Mhz & 100 m & 1 & 1 yr &0.5\\
    \hline
  \end{tabular}
     \end{center}
  \caption{Specifications of the CHIME-like experiment, see also Ref.~\cite{Hall:2012wd}.}
  \label{tab:chime}
\end{table}

The results for the two possible fiducial models described in the previous section are presented in Tab.~\ref{tab:chime12}, using $\ubotmin =
2\pi/\theta_{res}(z)$ with $\theta_{res}=\lambdasf (1+z)/D$.
Notice that the CHIME configuration can provide a high precision measurement of $w_0$. However, the precision 
in the extraction of the EDE background parameter $\omegae$, 
as well as  in the measurements of the dark energy clustering parameters $\cvis$ and $\ceff$, is quite poor. Concerning the standard cosmological parameters, the  constraints on both $\ln A_s$ and $\Omega_{\rm b} h^2$ are worse than those obtained with current CMB data. Indeed, these two parameters affect the overall amplitude of the 21 cm signal, while the CMB amplitude signal is mainly driven by the $\ln A_s$ parameter, with $\Omega_{\rm b} h^2$ controlling the CMB even-odd peak ratio. However, the constraints on both $\tau$ and $n_s$ are tighter for the 21 cm experiment. Let us emphasise that the addition of BICEP2 data does not change the results presented here.

\begin{table*}[t]
  \begin{tabular}{c|c|c c c}
 \hline
   &fiducial$_1$ (fiducial$_2$)&CHIME  &CHIME   \\
    & &  &+ Planck \& WP\\
\hline
$\Omega_{\rm b} h^2	$ & $ 	0.02258	$ & $ 	2.07\,(2.15) \cdot 10^{-3}	$ & $ 	2.55 \,(2.22)\cdot 10^{-4}	$ \\
$h	$ & $	0.71	$ & $	1.4\, (2.21)\cdot 10^{-2}	$ & $	0.88\,(1.11)\cdot 10^{-2}	$ \\
$\Omega_{\rm c} h^2	$ & $	0.1109	$ & $	7.07\, (9.57)\cdot 10^{-3}	$ & $	1.54\,(1.66)\cdot 10^{-3}	$\\
$\omegae	$ & $	0.01\, (0.03)	$ & $	1.92\, (2.97) \cdot 10^{-2}	$ & $	3.31\, (3.8)\cdot 10^{-3}	$ \\
$\cvis	$ & $	0.\, (0.33)	$ & $	13.6\,	(2.24)	$ & $	2.82\, (2.63)\cdot 10^{-1}	$ \\
$w_0	$ & $	-0.9	$ & $	5.35\, (7.78)\cdot 10^{-2}	$ & $	2.65\, (3.23)\cdot 10^{-2}	$ \\
$\ceff	$ & $	1.\, (0.33)	$ & $	0.214	\, (1.41)	$ & $	2.89 (2.72)\cdot 10^{-1}	$ \\
$n_s	$ & $	0.963	$ & $	1.9\, (3.63)\cdot 10^{-2}	$ & $	5.26\, (5.32)\cdot 10^{-3}	$ \\
$\tau	$ & $	0.088	$ & $	2.78\, (2.69)\cdot 10^{-3}	$ & $	7.14\, (6.79)\cdot 10^{-4}	$ \\
$\ln[10^{10} A_s]	$ & $	3.09784	$ & $	7.32\, (8.26)\cdot 10^{-1}	$ & $	2.44\, \, (2.44)\cdot 10^{-2}	$ \\
 \hline
  \end{tabular}
  \caption{1$\sigma$ errors on the parameters describing the two fiducial models here, which only differ in the values of the $\omegae$ and the dark energy clustering parameters.}
  \label{tab:chime12}
\end{table*}

\subsection{Omniscope $z>7$}
 \label{sec:om}
\begin{table}[h!]
  \begin{center}
  \begin{tabular}{c c c c c c c c}
    \hline
    redshift slices&$B_W$& $D$ & $f_{cover}$& $t_{obs}$ &$f_{sky}$\\
    \hline
9/10/11/12& 10 Mhz & 10 km & 0.1 & 1 yr &0.5\\
    \hline
  \end{tabular}
     \end{center}
  \caption{Specifications of the omniscope-like experiment for which we have considered $10^6$ antennas, see also Ref.~\cite{Clesse:2012th}.}
  \label{tab:omn}
\end{table}

We provide in Tab.~\ref{tab:omn}
the specifications of the future omniscope-like experiment explored
here. Table~\ref{tab:omn12} shows the $1\sigma$ errors for the
two fiducial models previously illustrated for the CHIME-like
experiment~\footnote{In this case, we have also marginalised over the
  parameter $\Delta_z$ specifying the duration of the reionisation process,
  see~\cite{Clesse:2012th} for more details on the background
  reionisation model.}.  While the $\Omega_{\rm b} h^2$ parameter can be measured 
  with a precision similar to the one achieved with current CMB
data, the $\ln A_s$ parameter is still better constrained by the
latter measurements. For the setup and the fiducial model considered here, the errors
on $\tau$ and $n_s$ are significantly better than for CMB experiments,
see also the discussion in Ref.~\cite{Clesse:2012th}. The addition of
Planck the and/or the BICEP2 priors does not change much the overall picture for
the marginalised errors depicted in Tab.~\ref{tab:omn}. Let us emphasise that we did not take into account extra ionising sources
that can severely damage the variances of the reionisation model
parameters, see e.g.~\cite{Mao:2008ug,Clesse:2012th}.

Concerning the dark energy parameters, the constraints on the
background parameters $w_0$ and $\omegae$ reach high precision levels, with 2\% and 7\% (6\%) errors, respectively, for 
$\omegae=0.01 (0.03)$. A similar precision on the measurement of 
the dark energy clustering parameters $\ceff$ and $\cvis$ is obtained with future 21 cm measurements, except for the case in which $c_{vis}^2= 0$
and $\ceff= 1.0$. For this particular scenario, the constraint on $\ceff$ is very poor. 

\begin{figure}[!]
   \begin{center}
     \hspace{-1cm}
     \begin{tabular}{c}
       \includegraphics[width=6cm]{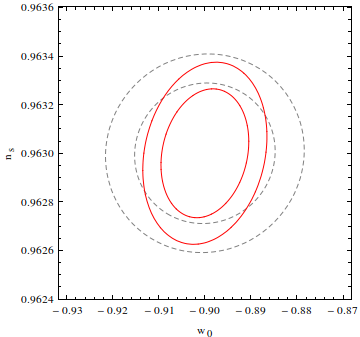} \\
       \includegraphics[width=6cm]{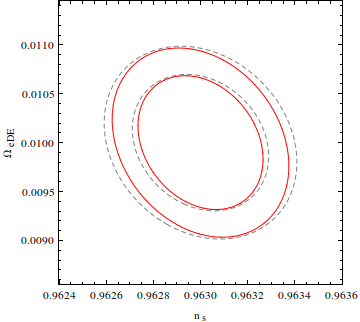}
       \end{tabular}
   \end{center}
   \caption{$1$ and $2\sigma$ allowed regions from the Fisher analysis with an
     omniscope-like experiment with four redshift slices
     $z=9,10,11,12$ for the fiducial model 1. The
     addition of the Planck measurements results in the continuous red
     contours.}
   \label{fig:omn12-2}
 \end{figure}

Figure~\ref{fig:omn12-2} shows the two-dimensional 1 and 2$\sigma$ allowed regions in a reduced number of parameters for the fiducial scenario with $\omegae= 0.01$,
$c_{vis}^2= 0$ and $\ceff= 1.0$.
The top panel of Fig.~\ref{fig:omn12-2} illustrates the expected
correlation in the $(w,n_s)$ plane. As in the case of the analysis of
Sec.~\ref{sec:current}, $\omegae$ and $w_0$ are anti-correlated, and therefore there exists a mild anti-correlation between $\omegae$
and $n_s$, as depicted in the bottom panel of
Fig.~\ref{fig:omn12-2}. We also depict in red solid lines the
resulting contours after adding the Planck measurements.

 \begin{table*}[t]
  \begin{tabular}{c|c|c c }
    \hline
    &fiducial$_1$ (fiducial$_2$)&Omniscope  &Omniscope \\
    & &  &+ Planck \& WP\\
    \hline
$\Omega_{\rm b} h^2$ & $ 	0.02258	$ & $ 	2.85\, (5.75)\cdot 10^{-5}	$ & $ 	2.64\,(4.79)\cdot 10^{-5}$\\
$h	$ & $	0.71	$ & $	5.51\, (5.54)\cdot 10^{-3}	$ & $	3.39\, (3.78)\cdot 10^{-3}$\\
$\Omega_{\rm c} h^2$	& $	0.1109	$ & $	2.51\, (5.73)\cdot 10^{-4}	$ & $	2.44\, (4.65)\cdot 10^{-4}$\\
$\omegae$ & $	0.01\, (0.03)	$ & $	0.697\, (1.6)\cdot 10^{-3}	$ & $	0.684\, (1.47)\cdot 10^{-3}$\\
$\cvis	$ & $	0.\, (0.33)	$ & $	1.93\, (1.4)\cdot 10^{-1}	$ & $	1.59\, (1.21)\cdot 10^{-1}$\\
$w_0	$ & $	-0.9	$ & $	1.53\, (1.56)\cdot 10^{-2}	$ & $	0.953 (1.09)\cdot 10^{-2}$\\
$\ceff	$ & $	1.\, (0.33)	$ & $	1.78\,	(0.22)	$ & $	2.86\, (1.7)\cdot 10^{-1}$\\
$n_s	$ & $	0.963	$ & $	2.89\, (4.27)\cdot 10^{-4}	$ & $	2.65\, (3.96)\cdot 10^{-4}$\\
$\tau	$ & $	0.088	$ & $	3.11\, (3.09)\cdot 10^{-5}	$ & $	3.1\, (3.08)\cdot 10^{-5}$\\
$\log[10^{10} A_s]	$ & $	3.09784	$ & $	3.34\, (3.18)\cdot 10^{-2}	$ & $	1.98\, (1.94)\cdot 10^{-2}$\\
$\Delta_z	$ & $	1.5	$ & $	8.39\, (8.8)\cdot 10^{-4}	$ & $	8.38\, (8.79)\cdot 10^{-4}$\\
    \hline
  \end{tabular}
  \caption{As Tab.~\ref{tab:chime12} but for the omniscope-like experiment considered here, see also Ref.~\cite{Clesse:2012th}.}
  \label{tab:omn12}
\end{table*}

\section{Summary and Conclusions}
\label{sec:conclusions}
In the last few years Cosmic Microwave Background (CMB) measurements have reached
an extremely high sensitivity, allowing for high precision cosmology
and providing, therefore, very tight constraints on the basic parameters governing the
standard $\Lambda$CDM model. The recent claimed detection of primordial B-modes from the BICEP2 experiment has also offered new insights in cosmology.
Here we have exploited the former signal, together with the latest CMB measurements, to update
the constraints on an Early Dark Energy component.  We find $\omegae<0.009$ at
$95\%$~CL when Planck, WMAP polarization and BICEP2 data are considered, assuming that the early dark energy component can be described by a perfect fluid. If the former assumption is relaxed, and the dark energy perturbation parameters $\ceff$ and $\cvis$ are allowed to vary freely, $\omegae$ turns out to be less well constrained.
Furthermore we find that current CMB measurements are unable to constrain $\ceff$ and $\cvis$.

In this case, future cosmological measurements of the $21$ cm line can be crucial. In the optimistic approach followed here (i.e. in the absence of  foregrounds or extra ionising sources), our Fisher matrix analyses of future data from an omniscope-like experiment show that the combination of these $21$ cm cosmological probes and current CMB measurements will be able to distinguish between the canonical quintessence scenario (characterised by $\ceff=1$ and $\cvis=0$) and other possible models  (with non standard clustering parameters, as, for instance, with $\ceff=0.33$ and $\cvis=0.33$) with $2\sigma$ significance, in the presence of a non-negligible early dark energy component $\omegae$. The errors on the energy density of the former parameter from the joint analysis of future $21$~cm data and current CMB measurements, assuming $\omegae\,=\,0.01 \, (0.03)$, are $0.684 \, (1.47)\cdot 10^{-3}$. Future $21$~cm probes can therefore achieve a precision below $10\%$ in the measurement of an early, non-homogeneous dark energy component.

\section*{Acknowledgements}
OM is supported by the Consolider Ingenio project CSD2007--00060, by
PROMETEO/2009/116, by the Spanish Grant FPA2011--29678 of the MINECO.
OM and MA are also partially supported by
PITN-GA-2011-289442-INVISIBLES.  We also thank the spanish MINECO
(Centro de excelencia Severo Ochoa Program) under grant SEV-2012-0249.
LLH is supported through an “FWO-Vlaanderen” post doctoral fellowship
project number 1271513. LLH also recognizes partial support from the
Strategic Research Program “High Energy Physics” of the Vrije
Universiteit Brussel and from the Belgian Federal Science Policy
through the Interuniversity Attraction Pole P7/37 “Fundamental
Interactions”.

\end{document}